\documentclass[aps,twocolumn,superscriptaddress]{revtex4-1}

\usepackage{graphicx,epsfig}
\usepackage{amsthm,dsfont,float,amsfonts,amsmath}

\begin{document}
\title{Seebeck effects in two-dimensional spin transistors}
\author{M. I. Alomar}
\affiliation{Institut de F\'{\i}sica Interdisciplin\`aria i Sistemes Complexos
IFISC (UIB-CSIC), E-07122 Palma de Mallorca, Spain}
\affiliation{Departament de F\'{\i}sica, Universitat de les Illes Balears, E-07122 Palma de Mallorca, Spain}
\author{Lloren\c{c} Serra}
\affiliation{Institut de F\'{\i}sica Interdisciplin\`aria i Sistemes Complexos
IFISC (UIB-CSIC), E-07122 Palma de Mallorca, Spain}
\affiliation{Departament de F\'{\i}sica, Universitat de les Illes Balears, E-07122 Palma de Mallorca, Spain}
\author{David S\'anchez}
\affiliation{Institut de F\'{\i}sica Interdisciplin\`aria i Sistemes Complexos
IFISC (UIB-CSIC), E-07122 Palma de Mallorca, Spain}
\affiliation{Departament de F\'{\i}sica, Universitat de les Illes Balears, E-07122 Palma de Mallorca, Spain}

\begin{abstract}
We consider a spin-orbit-coupled two-dimensional electron system under the influence of a thermal gradient externally applied to two attached reservoirs. We discuss the generated voltage bias (charge Seebeck effect), spin bias (spin Seebeck effect) and magnetization-dependent thermopower (magneto-Seebeck effect) in the ballistic regime of transport at linear response. We find that the charge thermopower is an oscillating function of both the spin-orbit strength and the quantum well width. We also observe that it is always negative for normal leads. We carefully compare the exact results for the linear response coefficients and a Sommerfeld approximation. When the contacts are ferromagnetic, we calculate the spin-resolved Seebeck coefficient for parallel and antiparallel magnetization configuration. Remarkably, the thermopower can change its sign by tuning the Fermi energy. This effect disappears when the Rashba coupling is absent. Additionally, we determine the magneto-Seebeck ratio, which shows dramatic changes in the presence of a the Rashba potential. 
\end{abstract}
\maketitle

\section{Introduction}\label{se:int}
One of the key ingredients of spintronics is the possibility of manipulating the electronic flow
via spin-charge coupling potentials \cite{fab07,joh85}. This goal can be achieved in semiconductor heterostructures
lacking space inversion symmetry, as demonstrated by Rashba \cite{ras60}.
Importantly, the strength of the Rashba spin-orbit interaction can be externally tuned with
a gate electrode electrostatically coupled to the 
heterostructure \cite{nit97,eng97,gru00}.
Thus, a spin field-effect transistor can be envisaged based
on a narrow channel sandwiched between two ferromagnetic electrodes with independently
controlled magnetizations \cite{dat90}.
The electric field applied to the gate is then viewed
as a momentum-dependent effective magnetic field acting on the channel,
causing a modulated precession of the injected spins, which are transmitted
into the drain electrode, provided their spin direction matches the drain magnetization.
This device works in the ballistic regime and in the single-mode limit, in which case
the precession angle is independent of energy.

Recent experiments report signatures of Rashba spin-orbit modulated conductance
in spin valve systems~\cite{koo09} measured in the nonlocal configuration \cite{jed02}.
An important difference
with the original proposal is that the transistor channel indeed supports multiple modes.
As a consequence, conductance oscillations disappear as the spin-orbit coupling increases,
a result which is consistent with numerical simulations \cite{gel10,zai11}.
The mechanism behind this amplitude decrease is the Rashba intersubband coupling
term that mixes adjacent subbands with opposite spin directions \cite{mir01}.
This term cannot be neglected in quantum wires and is responsible for a variety 
of phenomena: Fano-Rashba antiresonances \cite{san06,jeo06,zha06,per07,she08,san08,ore08,nag14},
removal of magnetic-field-induced
anomalous conductance steps \cite{ser05,qua10}, spin texturing \cite{hau01,gov02,she05,erl06,upa08,ste10,mal11,woj14},
and spin relaxation \cite{kis00,she05b,wu10}.

In the limit of wide channels, the spectrum of the transversal modes becomes continuous
and the Rashba precession and mixing terms must be treated on equal footing.
Reference \cite{gel11} shows that conductance oscillations persist, although they now 
originate from quantum interference between majority spin propagating states
and minority spin quasibound states. These two-dimensional electron systems (2DES)
offer the advantage of exhibiting transistor features in nonballistic devices \cite{sch03}
or even without the presence of magnetic contacts \cite{hal03}.
Quite recently, there is a renewed interest in the physics and applications of 2DES
due to their bandgap tunability and reduced dimensions \cite{fio14}.
 
Spintronic devices are attractive systems because they offer
new functionalities and less dissipation \cite{aws07}.
However, in contrast to the intense efforts in understanding the electric properties
of spin transistors, much less is known about their response to a temperature difference
(thermoelectric or Seebeck effect). The subject is interesting in view of recent
experiments that verify the generation of spin currents when a thermal gradient
is applied. This spin Seebeck effect has thus far been revealed in magnetic alloys \cite{uch08},
ferromagnetic semiconductors \cite{jaw10},
and even nonmagnetic materials \cite{jaw12}.
In a broad sense, spin caloritronics
is the study of spintronic effects created in thermally biased samples \cite{bau10}.
Spin biases due to temperature driven electron flow can be detected
by means of the inverse spin Hall effect \cite{sai06}.
Furthermore, a magneto-Seebeck
effect has also attracted a good deal of attention recently \cite{wal11,tei13,lop14,boe14}.
Here, the focus is not put on the creation of spin-polarized currents but rather
on the dramatic changes seen in the junction thermopower
when the orientation of the ferromagnetic electrodes is switched from
the parallel to the antiparallel configuration. Finally, recent measurements
of the charge thermopower in 2DES setups may indicate the presence of electronic correlations \cite{nar1,nar2}
or diffusive transport mechanisms \cite{chi09,gan12}.

Our aim in this work is to examine the three Seebeck effects (charge, spin and magneto)
in a 2DES spin transistor. We consider a quantum well laterally coupled to two
ferromagnetic reservoirs kept at different temperatures $\Delta T$, as sketched
in Fig.~\ref{fig:graph}.
Thermocurrent-induced spin polarization effects in Rashba 2DES
are discussed in Refs.\ \cite{ma10,wan10,dyr13,bor13,igl14}
using semiclassical approaches.
Here, we consider the quantum (ballistic) regime of transport and formulate
a theoretical model based on the scattering approach. We observe clear oscillations in
the charge Seebeck coefficient when the Rashba strength is varied even when
the 2DES is coupled to nonmagnetic reservoirs. Importantly, in this case the 
transmission for up spins and down spins are equal and hence the spin Seebeck
coefficient is identically zero. This result agrees with Ref.\ \cite{igl14},
which finds an absence of thermospin effects in spin-orbit coupled 2DES. 
However, we below show that spin imbalances can be created in response to temperature differences
when the electrodes are ferromagnetic. The effect is more prominent for larger
Rashba splittings.

Remarkably, the Seebeck coefficient
shows sign changes as a function of the Fermi energy for the case of magnetic leads.
This is one of the main findings  of our work, as it suggests the possibility of controlling
the thermoelectric current direction by varying the Fermi energy with a nearby gate.
Furthermore, the thermopower depends on the relative orientation of the leads'
magnetization. Since the spin-orbit field randomizes the spin direction of current-carrying
states, one would expect a quench of the magneto-Seebeck ratio when Rashba interaction
is turned on. However, we find significant changes in the presence of spin-orbit coupling.

\section{Theoretical model}\label{sec:mod}

\begin{figure}[t]
\centering
\includegraphics[width=0.45\textwidth]{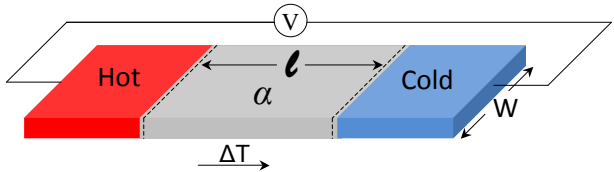}
\caption{(Color online) Pictorial representation of our system.
A two-dimensional semiconductor layer is formed inside a quantum well
subjected to a Rashba spin-orbit interaction of strength $\alpha$. The length of the central region where the Rashba coupling is constant is $\ell$ (gray area). Then, $\alpha$ decays smoothly to vanish in the contacts (colored areas). Electronic transport is induced with a bias voltage $V$ which is applied across the junction or with a temperature difference $\Delta T$ between the two contacts, hot and cold.}\label{fig:graph}
\end{figure}

We consider a two-dimensional semiconductor layer in the $xy$-plane with a central region of width $\ell$ subjected to a Rashba  spin-orbit interaction. In Fig.~\ref{fig:graph} we  show a sketch of our system. The Rashba coupling is spatially varying along the $x$-direction, which we take as the transport direction. The Hamiltonian reads
\begin{eqnarray}\label{eq:h}
\mathcal{H}=-\frac{\hbar^2}{2 m_0}\left(\frac{\partial^2}{\partial x^2}+\frac{\partial^2}{\partial y^2}\right)+\mathcal{H}_R\,,
\end{eqnarray}
where $m_0$ is the conduction-band effective mass of the electrons in 2DEG and $\mathcal{H}_R$ is the Rashba spin-orbit interaction depending on the Pauli matrices $\sigma_x$ and $\sigma_y$,
\begin{eqnarray}\label{eq:hr}
\mathcal{H}_R=-i\left(\alpha(x)\sigma_x\partial_y-\frac{1}{2}\sigma_y\left\lbrace \alpha(x),\partial_x\right\rbrace \right)\,.
\end{eqnarray}
In Eq.~\eqref{eq:hr} the Rashba strength $\alpha(x)$ takes a constant value in each of the three regions of our system. In the left and the right contacts $\alpha(x)$ is equal to zero while in the central region, of width $\ell$, $\alpha(x)$ takes a uniform value $\alpha(x)=\alpha$. In our numerical simulations, the variation of this parameter at the interfaces is almost abrupt with a minor numerical smoothing, see Appendix~\ref{app:f}. 

The eigenfunctions in the leads are plane waves since the reservoirs are assumed metallic with good screening properties. In those asymptotic regions the band structure takes the general form
\begin{eqnarray}\label{eq:k}
E=\frac{\hbar^2\mathcal{K}^2}{2m_0}=\frac{\hbar^2}{2m_0}\left(k^2+q^2\right)\,,
\end{eqnarray}
where $k$ and $q$ are the wave numbers along $x$ (longitudinal) and $y$ (transverse) directions, respectively. 

The $q$ momentum is constant throughout the system since 
the Hamiltonian of Eq.~\eqref{eq:h} remains invariant after translation along $y$.
Thus, the wave function in the central region can be written in terms of the product of
a plane wave in $y$-direction and an
$x$-dependent amplitude $\psi_{qs}(x)$ for each channel labeled with $(q,s)$.
If we sum over spins $s$ and over all the transverse momenta the total wave function reads
\begin{eqnarray}\label{eq:f}
\Psi(x,y,\eta)=\sum_{s=\pm}\int dq \,\psi_{qs}(x) e^{iqy}\chi_s(\eta)\,,
\end{eqnarray}
where $s$ is a spin index that characterizes the spin function 
$\chi_s(\eta)\equiv\langle\eta|s\rangle$, with $\eta=\uparrow,\downarrow$
the usual basis of the Pauli matrices. 

To determine the channel amplitude equations we project the Schr\"odinger equation
$(\mathcal{H}-E)\Psi=0$ on a particular channel $(q,s)$, 
\begin{eqnarray}\label{eq:eq}
&&\left(-\frac{\hbar^2}{2 m_0}\frac{d^2}{dx^2}+\frac{\hbar^2q^2}{2 m_0}-E\right)\psi_{qs}(x)\\ \nonumber
&+&\sum_{s^{\prime}=\pm}\left\lbrace\left(\alpha(x)q\left\langle s\vert\sigma_x\vert s^{\prime}\right\rangle+\frac{i}{2}\alpha^{\prime}(x)\left\langle s\vert \sigma_y\vert s^{\prime}\right\rangle\right)\psi_{qs^{\prime}}(x)\right.\\\nonumber
&&\left.\quad\quad\quad +\,i \,\alpha(x)\left\langle s\vert \sigma_y\vert s^{\prime}\right\rangle\frac{d}{dx}\psi_{qs^{\prime}}(x)\right\rbrace=0\,.
\end{eqnarray}
These channel equations show coupling between different spins due to the spin-orbit interaction (the terms $\left\langle s\vert \sigma_{x,y}\vert s^{\prime}\right\rangle$). However, channels with different $q$ remain uncoupled due to the translational invariance along $y$-direction. This a unique feature of 2DES devices since in quantum wires the coupled channel method
does connect adjacent modes with opposite spins \cite{san06}.

In the contacts the Rashba coupling vanishes and the wave functions can be expressed with the aid of
input $a^{(c)}_{qs}$ and output $b^{(c)}_{qs}$ amplitudes,
\begin{equation}
\label{bcond}
\psi^{(c)}_{qs}(x)=a^{(c)}_{qs} e^{i s_c kx}+b^{(c)}_{qs} e^{-is_c kx}\; ,
\end{equation}
where $c=L$ ($c=R$) for the left (right) lead and we take $s_L=+$ and $s_R=-$. 
Within scattering theory the output amplitudes are determined from the input ones
via reflection $r^{(c)}_{s's}$ and transmission $t^{(c)}_{s's}$ 
amplitudes,
\begin{equation}
\label{smat}
b^{(c)}_{qs'} = 
\sum_{s}{
r^{(c)}_{s's}\, a^{(c)}_{qs}
+
t^{(c)}_{s's}\, a^{(\bar{c})}_{qs}
}\;.
\end{equation}
Here, $\bar{c}$ denotes the opposite contact to $c$. We note that the matrices $t^{(c)}$ and $r^{(c)}$ of Eq.~\eqref{smat} depend on $q$ 
and they are found after discretizing Eq.~\eqref{eq:eq} on a grid and imposing the boundary conditions given by Eq.~\eqref{bcond}
with the quantum-transmitting-boundary algorithm \cite{len90}.

\begin{figure}[t]
\centering
\includegraphics[width=0.45\textwidth]{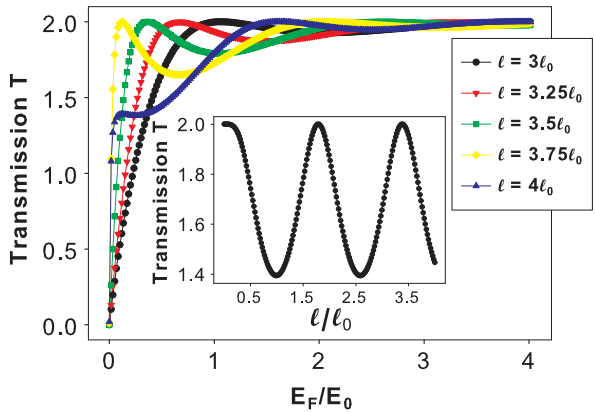}
\caption{(Color online) Total transmission probability as a function of the Fermi energy for transverse momentum $q=0$, Rashba coupling  $\alpha=1.7 \alpha_0$ and different lengths of the central region $\ell$'s. Inset: Total transmission probability as a function of $\ell$ for $q=0$, $\alpha=1.7 \alpha_0$ and $E_F=0.5E_0$. We take $\ell_0=60\,$nm. Hence, $E_0=\hbar^2/(m_0 \ell_0^2)=0.92\,$meV and $\alpha_0=\hbar^2/(m_0 \ell_0)=55.2\,$meV~nm for $m_0=0.023m_e$ as in InAs.}\label{fig:tl}
\end{figure}

We define the transmission probabilities
\begin{equation}
T_{s^{\prime}s}(q,E) = \left|t^{(R)}_{s's}\rule{0cm}{0cm}\right|^2\; ,\quad  
T'_{s^{\prime}s}(q,E) = \left|t^{(L)}_{s's}\rule{0cm}{0cm}\right|^2\; ,
\end{equation}
where $T_{s^{\prime}s}(q,E)$ represents the probability that an incident electron in the left lead with spin $s$ is transmitted to the right lead with spin $s^{\prime}$. Analogously, $T^{\prime}_{s^{\prime}s}(q,E)$ is the transmission probability for an electron injected from the right contact with spin $s$ to arrive at the left contact with spin $s^\prime$. 
Figure~\ref{fig:tl} shows the total transmission, $T=\sum_{ss^{\prime}}T_{s^{\prime}s}$, as a function of the Fermi energy for the case $q=0$ at fixed $\alpha$ and different lengths of the Rashba region. The $q=0$ mode is interesting since it corresponds to normal incidence, thus dominating the total transmission.

We observe in Fig.~\ref{fig:tl} that $T$ quickly reaches a maximum and then oscillates. This behavior can be nicely understood from the one-dimensional problem of electrons scattering off a square well where $\ell$ is its width and $\alpha$ is proportional to its depth. To see this, we set $q=0$ in Eq.~\eqref{eq:eq},
\begin{eqnarray}\label{eq:eqq0}
\left(-\frac{\hbar^2}{2 m_0}\frac{d^2}{dx^2}-E\right)\psi_{0s}(x)&+& \frac{i}{2}\alpha^{\prime}(x)\, s \,\psi_{0s}(x)\\ \nonumber
&+&i\, \alpha(x)\,s\,\frac{d}{dx}\psi_{0s}(x)=0\,,
\end{eqnarray}
where we choose the quantization axis of the spin $s=\pm$ in the $y$-direction. Clearly, the two spins are uncoupled. We now make the gauge transformation
\begin{eqnarray}
\psi_{qs}(x)&=&e^{i s\frac{m_0}{\hbar}\int^x\alpha(x^{\prime}) dx^{\prime}}\widetilde{\psi}_{qs}(x)\,.
\end{eqnarray}
Then Eq.~\eqref{eq:eqq0} is written in terms of $\widetilde{\psi}_{qs}(x)$
\begin{eqnarray}\label{eq:pozo}
-\frac{\hbar^2}{2 m_0}\widetilde{\psi}_{0s}^{\prime\prime}(x)+\left(V_0(x)-E\right)\widetilde{\psi}_{0s}(x)=0
\end{eqnarray}
where $V_0(x)=-m_0\alpha^2(x)/2\hbar^2$. For a piecewise constant $\alpha(x)$, Eq.~\eqref{eq:pozo} corresponds to the Hamiltonian of a square well of depth $V_0={m_0 \alpha^2}/{(2 \hbar^2)}$. We can readily write the total transmission
(summed over spins) as \cite{griffiths}
\begin{eqnarray}\label{eq:t}
T=\frac{2}{1+\frac{V_0^2}{4 E_F(V_0+E_F)}\sin^2 \left(\ell \frac{\sqrt{2 m_0(E_F+V_0)}}{\hbar}\right)}\,,
\end{eqnarray}
where $\ell$ is the width of the square well (the spin-orbit region length in our case). Then, $T$ is maximum whenever
\begin{eqnarray}\label{eq:rel1}
\frac{\ell}{\hbar} \sqrt{2 m_0(E_F+\frac{m_0 \alpha^2}{2 \hbar^2})}=n \pi\,,
\end{eqnarray}
where $n=0,1,2,\ldots$. The first maximum of $T$ for $E_F>0$ corresponds to $n=1$.
As we increase the width of the spin-orbit region the first maximum shifts to lower energy, in perfect agreement
with Fig.~\ref{fig:tl}. The maximum displacement occurs because $E_F$ should decrease to maintain the condition of Eq.~\eqref{eq:rel1} as $\ell$ increases.  In the inset of Fig.~\ref{fig:tl}, we represent the total transmission as a function of $\ell$ for fixed values of $\alpha$  and $E_F$. $T$ presents maxima at equidistant values of $\ell$. This can also be simply understood from the resonant condition derived from Eq.~\eqref{eq:rel1}.

\begin{figure}[t]
\centering
\includegraphics[width=0.45\textwidth]{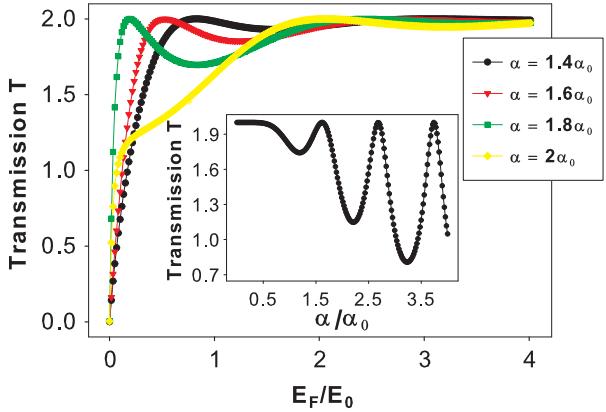}
\caption{(Color online) Total transmission probability as a function of the Fermi energy for $q=0$, $\ell=3.5\ell_0$ and different values of the Rashba strength $\alpha$. Inset: Total transmission probability as a function of $\alpha$ for $q=0$, $\ell=3.5\ell_0$ and $E_F=0.5E_0$. }\label{fig:ta}
\end{figure}

In Fig.~\ref{fig:ta} we plot $T$ as a function of the Fermi energy for $q=0$ at fixed length $\ell$ and different Rashba magnitudes. Similarly to Fig.~\ref{fig:tl}, as $\alpha$ increases the first maximum shifts to lower energies. In this case, the analogy with the problem of the square well is based on $\alpha$, which is proportional to the potential depth. Thus, as we increase the value of $\alpha$, the Fermi energy should decrease to fulfil the maxima condition of Eq.~\eqref{eq:rel1}. The inset of Fig.~\ref{fig:ta} shows the $T$ as a function of $\alpha$ for fixed $E_F$ and $\ell$. Notice that the spacing between maxima increases smoothly because the resonant condition
\begin{eqnarray}\label{eq:rel3}
\alpha_{\text{max}}=\sqrt{\frac{2 \hbar^2}{m_0}\left(\frac{1}{2m_0}\left(\frac{n \pi \hbar}{\ell}\right)^2-E_F\right)}
\end{eqnarray}
shows a nonlinear dependence with $n$.

\begin{figure}[t]
\centering
\includegraphics[width=0.45\textwidth]{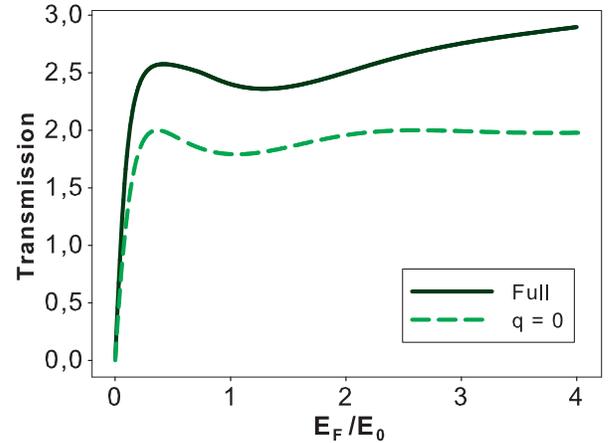}
\caption{(Color online) Total transmission probability as a function of the Fermi energy for $q=0$ (dashed curve) and integred over all $q$'s (solid curve). Parameters: $\ell=3.5\ell_0$ and $\alpha=1.7\alpha_0$.}\label{fig:tfq0p}
\end{figure}

The transmission probabilities for $q\neq0$ cannot be expressed in closed analytical form because the spins become coupled as illustrated in Eq.~\eqref{eq:eq}. It turns out that the contribution of these modes to the total transmission for a given energy is small as electrons are more likely to be refracted from the junction interfaces if $q\neq 0$.  To verify this,
let $\mathcal{T}_s$ be the spin-transmission summed over all transverse momenta and transmitted spins. 
Since we consider transport along the longitudinal ($x$) direction,
we include the projection factor $\cos\theta$,
\begin{eqnarray}\label{eq:ts}
\mathcal{T}_s(E)=2\sum_{s^{\prime}}\int_{0}^{\pi/2}\!\!\!\!T_{s^{\prime}s}(E,\theta)\cos \theta d\theta\,.
\end{eqnarray}
The total projected transmission is thus $\mathcal{T}=\sum_s\mathcal{T}_s$.
In Fig.~\ref{fig:tfq0p} we compare $\mathcal{T}$ from Eq.\ \eqref{eq:ts} with its $q=0$ contribution.
Clearly, the $q=0$ mode contributes more than $75\%$ to the full transmission for most energies. This 
confirms that with the mode $q=0$ we can understand the basic dynamics of the transmission through our system.
For completeness, however, in our calculations we take into account all channels to carefully assess the transport responses.

\section{Current transport}\label{sec:cur}

Within the scattering approach for mesoscopic transport, the current 
$I_s$
in the $x$ direction for a given spin $s$ 
can be written in terms of the transmission found in Sec.~\ref{sec:mod} (see Appendix~\ref{app:i}):
\begin{eqnarray}\label{eq:Is}
I_s=\frac{e}{h}\frac{W}{2 \pi}\int_{0}^{\infty}\!dE\,
\mathcal{T}_s(E)\,\mathcal{K}(E)\,
\left[\rule{0cm}{0.4cm}
f_{Ls}(E)-f_{Rs}(E)\right]\;,\nonumber\\
\end{eqnarray}
where $W$ is the length in the transverse direction and $\mathcal{K}(E)$ is the momentum in the contacts derived from Eq.~\eqref{eq:k}. 
Here $f_{cs}(E)$ with $c=L,R$
are the Fermi-Dirac distribution functions of electrons in each contact with spin $s$,
\begin{equation} \label{eq:fermi}
f_{cs}(E) = f(E-\mu_{cs},T_c) \equiv \frac{1}{1+e^{(E-\mu_{cs})/k_BT_c}}\; ,
\end{equation}
with
the electrochemical potential $\mu_{cs}$ and the lead temperature $T_c$ 
specified below. In cases of common background temperature we denote this
by $T_0$ while if no biases are applied $\mu_{cs}=E_F$.
The total current is, finally, $I=\sum_sI_s$.

From the electrochemical potentials $\mu_{cs}=E_F+eV_{cs}$ ($V_{cs}$ is the voltage in contact $c$ for electrons with spin $s$
which accounts for possible  population imbalances between different spin subbands)
and contact temperatures $T_c$, we define the temperature difference $\Delta T$, the mean electrochemical potential $\mu_c$, the bias voltage $V$ and  the spin bias $V_{\mathcal{S}}$ using the following relations:
\begin{subequations}\label{eq:def}
\begin{eqnarray}
&\Delta T&= T_L-T_R\,,\\
&\mu_{c}&=\left(\mu_{c+}+\mu_{c-}\right)/2\,,\\
&eV&=\mu_L-\mu_R\,,\\
&eV_{\mathcal{S}}&=\left(\mu_{L+}-\mu_{L-}\right)-\left(\mu_{R+}-\mu_{R-}\right)\,.
\end{eqnarray}
\end{subequations}

We focus now on the response of our 2D spin transistor to small values
of $V$, $\Delta T$ and  $V_{\mathcal{S}}$. After a Taylor expansion of  Eq.~\eqref{eq:Is} up to first order in these shifts, we find
\begin{eqnarray}\label{eq:I1}
I_s=G_s \frac{\mu_{Ls}-\mu_{Rs}}{e} +L_s\, \Delta T \,,
\end{eqnarray}
where the transport coefficients $G_s$ and $L_s$ are the
linear electric and thermoelectric conductances, respectively. 
Using Eqs.~\eqref{eq:def} in Eq.~\eqref{eq:I1}, we obtain the total current
\begin{eqnarray}
\label{eq:Itot}
I&=&\left(G_++G_-\right)V+\frac{1}{2}\left(G_+-G_-\right)V_{\mathcal{S}}\\ 
\nonumber
&&+\left(L_++L_-\right)\Delta T\,,
\end{eqnarray}
and the spin current, defined as $I_{\mathcal{S}}\equiv I_+-I_-$,
\begin{eqnarray}\label{eq:Ispin}
I_{\mathcal{S}}&=&\left(G_+-G_-\right)V+\frac{1}{2}\left(G_++G_-\right)V_{\mathcal{S}}\\ \nonumber
&&+\left(L_+-L_-\right)\Delta T\,.
\end{eqnarray}
Both current expressions are valid in linear response.

\begin{figure}[t]
\centering
\includegraphics[width=0.45\textwidth]{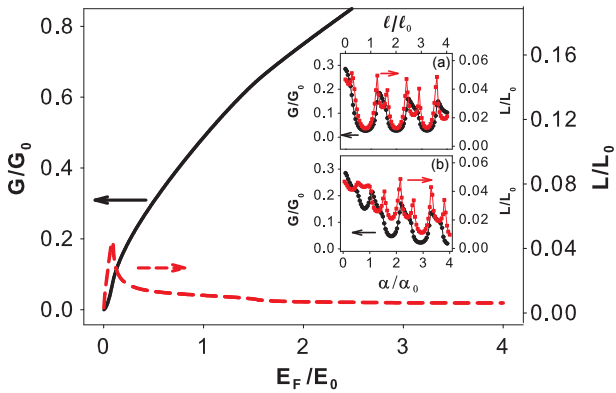}
\caption{ (Color online) Electric conductance (left axis) and thermoelectric conductance (right axis) as a function of the Fermi energy at low temperature. Parameters: $k_BT_0=0.01E_0$, $\ell=2.9\ell_0$ and $\alpha=2.6\alpha_0$. The electric conductance as a function of $\ell$($\alpha$) is plotted in the left axis of the inset (a)[(b)]. In the right axis of the same inset the thermoelectric conductance is depicted. Parameters of inset (a)[(b)]: $k_BT_0=0.01E_0$, $E_F=0.1E_0$ and $\alpha=2.6\alpha_0$($\ell=2.9\ell_0$).}\label{fig:gllap}
\end{figure}

In Eq.~\eqref{eq:I1} the electric conductance is given by 
\begin{eqnarray}\label{eq:g}
\frac{G_s}{G_0}=\frac{1}{2\pi}\!\int_{0}^{\infty}dE\,\,\mathcal{K}(E) \left(-\frac{\partial f}{\partial E}\right)\mathcal{T}_s(E)\,,
\end{eqnarray}
where $G_0=e^2W/h$ is the unit of conductance for a 2DES length $W$
along $y$. Physically, $G_s/G_0$ is thus a 
spin-resolved conductance per unit
of transverse length.
Hence, the total conductance is $G=\sum_sG_s$.
In the zero temperature limit, the conductance per spin reads
\begin{eqnarray}\label{eq:gt0}
\frac{G_s}{G_0}=\frac{\mathcal{K}_F}{2\pi}\,\mathcal{T}_s(E_F)\equiv g_s(E_F)\,,
\end{eqnarray}
where $\mathcal{K}_F=\mathcal{K}(E_F)=\sqrt{2 m_0 E_F}/\hbar$. 

The linear thermoelectric conductance in Eq.~\eqref{eq:I1} reads
\begin{eqnarray}
\label{eq:l}
\frac{L_s}{L_0} &=& \frac{1}{k_B T_0 2\pi} 
\int_{0}^{\infty}dE\,\mathcal{K}(E) \!\left(E\!-\!E_F\right)\!\left(\!-\frac{\partial f}{\partial E}\right)\!\mathcal{T}_s(E)\,,\nonumber\\
\end{eqnarray}
where $L_0=e k_B W/h$ is the natural scale of the thermoelectric response of a 2DES of width $W$.
The total thermoelectric conductance is $L=\sum_s L_s$. At very low temperatures, a Sommerfeld expansion~\cite{Ashcroft}
yields
\begin{equation}\label{eq:lt0}
\frac{L_s}{L_0}=\frac{k_B T_0 \pi}{6E_F}\mathcal{K}_F\left(\frac{1}{2}\mathcal{T}_s(E_F)+E_F\frac{\partial \mathcal{T}_s}{\partial E_F}\right)\,,
\end{equation}
where the derivative is defined as $\frac{\partial}{\partial E_F}\equiv\left.\frac{\partial}{\partial E}\right|_{E=E_F}$.

Figure~\ref{fig:gllap} presents the total electric and thermoelectric conductances as a function of the Fermi energy from a calculation of Eqs.~\eqref{eq:gt0} and~\eqref{eq:lt0}, respectively. The transmission is determined from a full evaluation of Eq.~\eqref{eq:ts}. We find that the electric conductance (black solid line) is an monotonically increasing function of $E_F$. At sufficiently large energy values,
$G$ goes as $E_F^{1/2}$ because the transmission approaches its maximum value when $E_F\gg E_0$ (see Figs.~\ref{fig:tl} and~\ref{fig:ta}). Only at low energies $G$ presents a small deviation from this behavior, which is much more visible when we calculate the thermoelectric conductance (red dashed line). 
While at high energies $L$ approaches zero as $E_F^{-1/2}$ as dictated by Eq.~\eqref{eq:lt0},
the thermoelectric conductance shows a peak at low values of $E_F$.

The insets of Fig.~\ref{fig:gllap} depict $G$ and $L$ as a function of the spin-orbit region width (top panel) and the Rashba coupling intensity (bottom panel). We observe that both responses are strongly modulated and present maxima at fixed values. This originates from the transmission oscillations discussed in Sec.~\ref{sec:mod} as a function of $\alpha$ or $\ell$. Interestingly, our device works as a current modulator in response to voltage or temperature biases applied to normal leads. This is in contrast with the spin transistor proposal of Ref.\ \cite{dat90}.
Our system (both the leads and the 2DES channel) is entirely nonmagnetic. It is precisely due to this reason that $G_+=G_-$ and $L_+=L_-$ and no spin polarization is possible.

\section{Charge and spin thermopower}\label{sec:ther}

In Eq.~\eqref{eq:lt0} we find that the thermoelectric conductance comprises
two terms at low temperature, in contrast to the electric conductance, Eq.~\eqref{eq:gt0},
which consists of a single term. With the aim to understand more clearly this additional contribution we calculate the charge thermopower or Seebeck coefficient, $S=(V/\Delta T)_{I=0 ,I_{\mathcal{S}}=0}$. $S$ determines the voltage generated in the junction in response to a temperature shift under open circuit conditions. Then, from Eqs.~\eqref{eq:Itot} and~\eqref{eq:Ispin}, we find
\begin{eqnarray}\label{eq:s}
S=-\frac{1}{2}\left(\frac{L_+}{G_+}+\frac{L_-}{G_-}\right)\,.
\end{eqnarray}

\begin{figure}[t]
\centering
\includegraphics[width=0.45\textwidth]{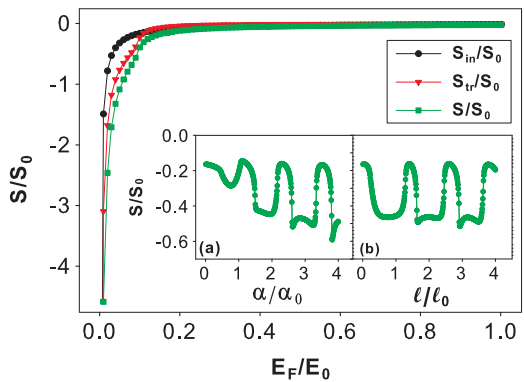}
\caption{(Color online) Seebeck coefficient at low temperature as a function of the Fermi energy. Parameters: $k_BT_0=0.01E_0$, $\ell=2.9\ell_0$ and $\alpha=2.6\alpha_0$. The two insets, (a) and (b), show $S$ as a function of $\alpha$ and $\ell$, respectively. Parameters of inset (a)[(b)]: $k_BT_0=0.01E_0$, $E_F=0.1E_0$ and $\ell=2.9\ell_0$($\alpha=2.6\alpha_0$).}\label{fig:slap}
\end{figure}

Inserting Eqs.~\eqref{eq:gt0} and~\eqref{eq:lt0} in Eq.~\eqref{eq:s}, we obtain the low-temperature thermopower,
\begin{eqnarray}\label{eq:sexp}
\frac{S}{S_0}=- \frac{k_BT_0\pi^2}{3 E_F}\left(\frac{1}{2}+E_F\sum_s\frac{\partial\mathcal{T}_s/\partial E_F}
{\mathcal{T}_s(E_F)}\right)\,,
\end{eqnarray}
The Seebeck coefficient is measured in units of $S_0=k_B/e$.
We observe two contributions in Eq.~\eqref{eq:sexp}.
The first term $S_{in}=-\pi^2 k_B T_0/6E_F$ is constant and represents an intrinsic contribution to the entropy per unit charge of
thermally excited electrons. It is independent of the scattering potential and the sample details.
The second term, $S_{tr}=-(\pi^2 k_B T_0/3)\sum_s({\partial\mathcal{T}_s/\partial E_F})/{\mathcal{T}_s}$,
is a purely transport contribution that arises from the energy dependence of the transmission function.
Interestingly, the first term is unique to 2DES since in quasi-onedimensional systems (quantum point contacts)
and quasi-zerodimensional systems (quantum dots) this contribution is absent.
Therefore, the intrinsic term can be explained as a dimensionality effect which also shows up in, e.g., graphene
monolayers~\cite{alo14}. The only difference is that in graphene $S_{in}$ is doubled because
its energy dispersion is linear, $E\sim\mathcal{K}$, unlike the quadratic dependence in our
semiconductor 2DES [Eq.~\eqref{eq:k}]. In fact,
if we use the $g_s$ function defined in Eqs.~\eqref{eq:gt0} we can recast Eq.~\eqref{eq:sexp}
as a Mott-like formula,
\begin{eqnarray}\label{eq:sg}
\frac{S}{S_0}=-\frac{k_BT_0\pi^2}{3}\sum_s\frac{\partial \ln g_s(E_F)}{\partial E_F}\,.
\end{eqnarray} 
The difference with the formula discussed in Ref.~\cite{mott} is that our Eq.~\eqref{eq:sg}
is valid in the ballistic regime of quantum transport.

\begin{figure}[t]
\centering
\includegraphics[width=0.45\textwidth]{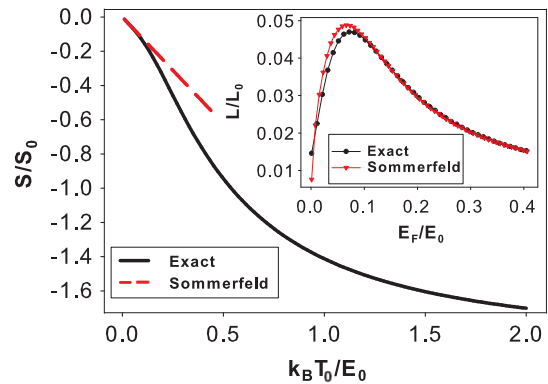}
\caption{(Color online) Thermopower as a function of the base temperature for both the exact calculation and the lowest order Somerfeld approximation. Parameters: $E_F=E_0$, $\ell=3.5\ell_0$ and $\alpha=1.7\alpha_0$. The inset presents a comparison between the exact calculation of the thermoelectric conductance and the Sommerfeld result as a function of the Fermi energy. Parameters of the inset: $k_BT_0=0.01E_0$, $\ell=3.5\ell_0$ and $\alpha=1.7\alpha_0$.}\label{fig:ssnsp}
\end{figure}

We analyze the relative importance between the intrinsic and the transport terms. To visualize this,
Fig.~\ref{fig:slap} shows the different contributions as a function of the Fermi energy. In general,
both have comparable strengths and should then be treated on equal footing.
 Note that $S$ is always negative because the generated voltage tends to counteract the thermal bias.
 Below, we will show deviations of this behavior when the sample is attached to ferromagnetic contacts. In Fig.~\ref{fig:slap} $S_{in}$ grows as $1/E_F$ and quickly goes to zero. On the other hand, although the overall trend of $S_{tr}$ is similar to the intrinsic term, we observe in $S_{tr}$ a stronger effect due to the potential scattering at low energy. Finally, we recall that both $G$ and $L$ present oscillations as a function of $\ell$ and $\alpha$, see insets (a) and (b) of Fig.~\ref{fig:gllap}. The thermopower, which is the ratio between the thermoelectric and electric conductances, also presents these oscillations as seen in the insets (a) and (b) of Fig.~\ref{fig:slap}. Importantly, the oscillations are due to the transport term since $S_{in}$ is independent of $\alpha$ and $\ell$.

We note in passing that the main properties of $S$ discussed above can be also detectable in heat current measurements. Due to Onsager reciprocity, the electrothermal conductance (Peltier effect) is directly connected to the thermoelectric conductance (Seebeck effect),
see App.~\ref{app:q}. Moreover, we show in App.~\ref{app:q} that the Wiedemann-Franz law holds at low temperature. Then, it is natural to ask to what extent the Sommerfeld approximation is valid. To do so, we compare the results of Eqs.~\eqref{eq:g} and~\eqref{eq:l}, i.e., the exact calculation at $k_BT_0=0.01E_0$, with   Eqs.~\eqref{eq:gt0} and~\eqref{eq:lt0}, which are valid at very low temperature. The difference between the exact and Sommerfeld calculation of the thermoelectric conductance is shown in the inset of Fig.~\ref{fig:ssnsp}. At low energies, $E_F/E_0<0.1$, there exists a small deviation between the exact curve and the Sommerfeld result. This confirms that the Sommerfeld expansion is no longer valid when $E_F\sim k_BT_0$. The main panel of Fig.~\ref{fig:ssnsp} presents the deviation of the Sommerfeld calculation (dashed curve) for the Seebeck coefficient compared with the exact  value (solid curve) as a function of the background temperature. In the Sommerfeld expansion, $S$ depends linearly with $T_0$ as shown in Eq.~\eqref{eq:sexp}. Therefore, for high temperatures such that $k_BT_0/E_F>0.1$ the Sommerfeld approximation is not valid. 
 
We end this section considering
the spin Seebeck coefficient $S_{\mathcal{S}}$. When we apply a temperature gradient across the junction, it can also lead to spin accumulations in the leads. Then, the spin Seebeck coefficient $S_{\mathcal{S}}=(V_{\mathcal{S}}/\Delta T)_{I=0,I_{\mathcal{S}}=0}$ determines the spin voltage $V_{\mathcal{S}}$ generated in the junction in response to a temperature shift when both the charge and the spin currents are set to zero. From Eqs.~\eqref{eq:Itot} and~\eqref{eq:Ispin} we obtain
\begin{eqnarray}\label{eq:ss}
S_{\mathcal{S}}=-\left(\frac{L_+}{G_+}-\frac{L_-}{G_-}\right)\,.
\end{eqnarray}
For normal leads the spin Seebeck is identically zero since the system is nonmagnetic and the spin polarization is not possible. 
We consider in the next section a 2DES coupled to ferromagnetic leads, a system where $S_{\mathcal{S}}$ does not necessarily vanish.

\section{Ferromagnetic contacts}\label{sec:fer}

The case of ferromagnetic contacts is relevant for spin-injection problems in ferromagnet-semiconductor junctions \cite{fab07}.
We describe the contacts with the Stoner-Wohlfarth model for itinerant ferromagnetism. The electronic bands for opposite spins become split due to exchange interaction between carriers. Then, the Hamitonian reads
\begin{eqnarray}\label{eq:hd}
\mathcal{H}_{\Delta}= \Delta(x)\,\hat{n}\,\cdot\,\vec{\sigma}+|\Delta(x)|\,,
\end{eqnarray}
where $\hat{n}$ is the magnetization direction in which the leads are polarized, $\vec\sigma$ is the spin vector and $\Delta(x)$ is the Stoner splitting, which is finite in the $x$-positions of the leads only. For metallic electrodes, $\Delta$ is of the order of a few eV, introducing serious conductivity mismaches at the junction \cite{sch00}, a problem which can be mitigated using tunnel barriers \cite{ras00}. For the present discussion, it is more convenient to consider spin injectors made of diluted magnetic semiconductor compounds, which show giant Zeeman splittings $\Delta$ up to $20$~meV for moderate values of external magnetic fields \cite{slob}. For definiteness, we consider in Eq.~\eqref{eq:hd} an energy shift $|\Delta(x)|$ that eases the spin transport analysis. In particular, this shift eliminates the effect of a potential mismatch and can be experimentally implemented with a potential gating of the central regions.

\begin{figure}[t]
\centering
\includegraphics[width=0.44\textwidth]{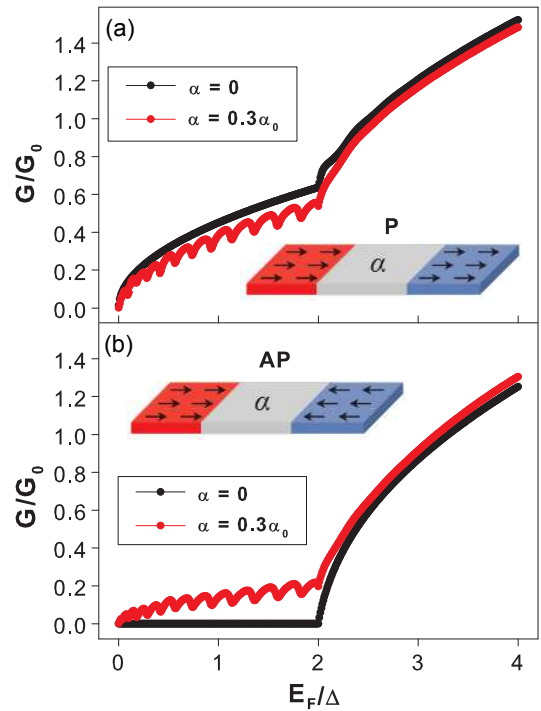}
\caption{(Color online) Electric conductance as a function of the Femi energy for P (a) and AP (b) polarizations along the $x$-direction. Red and black curves represent the case with or without spin-orbit interaction in the central region, respectively. 
The Stoner field splitting 
$\Delta=10\,$meV is taken as the energy unity. This way $\ell_0=\sqrt{\hbar^2/m_0\Delta}=18.2\,$nm and $\alpha_0=\sqrt{\hbar^2\Delta / m_0}=182\,$meV nm. Parameter: $\ell=8\ell_0$.}\label{fig:gp}
\end{figure}

In the same way of the piecewise constant function $\alpha(x)$ in Sec.~\ref{sec:mod}, we take a uniform $\Delta(x)$ in each of the three parts of our system: nonzero $\Delta_c$ in the contacts and vanishingly small $\Delta$ in the 2DES. For the numerical implementations, the interfaces are described with slightly smoothed functions, see App.~\ref{app:f}.
We investigate two different orientation schemes: parallel (P) and antiparallel (AP) configurations as sketched in the insets of Fig.~\ref{fig:gp}. The case of parallel polarization corresponds to $\Delta_L=\Delta_R\equiv\Delta$ whereas the antiparallel case results from $\Delta_L=-\Delta_R\equiv\Delta$, where $\Delta$ is half of the absolute Zeeman splitting. In what follows, we take the polarization in the $x$-direction: $\hat{n}\,\cdot\,\vec{\sigma}=\sigma_x$. Thus, we define a potential in the leads,
\begin{eqnarray}\label{eq:vs}
v_s(x)=s\Delta(x)+|\Delta(x)|\,,
\end{eqnarray}
where $s=\pm$ labels the spin in the $x$-direction. Then, for the P configuration the electrons with $s=+$ (minority spins) are confined by a potential well of width $d$ (the separation between contacts, see App.~\ref{app:f}) while those with $s=-$ feel no potential as they travel (majority spins). In the AP configuration both types of carrier experience a potential step but localized in opposite contacts. 
Due to the differing potential landscapes, we expect strong changes as compared with the nonmagnetic case treated
in the previous sections.

The dispersion relation on the contacts generalizes Eq.~\eqref{eq:k} conveniently modified to account for the Stoner field:
\begin{eqnarray}\label{eq:e2}
E=\frac{\hbar^2 \mathcal{K}^2_{cs}}{2m_0}+s\Delta_c+|\Delta_c|\,,
\end{eqnarray}
Importantly, the total momentum $\mathcal{K}_{cs}$ now depends on both the contact and the spin. As a consequence,
since the $q$-momentum remains invariant, we find 
\begin{eqnarray}\label{eq:k2}
\mathcal{K}_{cs}^2=k^2_{cs}+q^2=\frac{2m_0}{\hbar^2}\left(E-s\Delta_c-|\Delta_c|\right)\,.
\end{eqnarray}
The channel amplitude equation expressed in Eq.~\eqref{eq:eq} keeps the same form,
with the only addition of the spin dependent potential 
$v_s(x)$ and the replacement $k\to k_{cs}$ in the asymptotic 
conditions Eq.\ (\ref{bcond}).
From the resulting expression we calculate the new probability transmissions and obtain, finally, the total current in the propagation direction,
\begin{eqnarray}\label{eq:Is2}
I_s&=& \frac{e}{h}\frac{W}{2\pi}\int_{0}^{\infty} dE \,\mathcal{T}_s(E)\,\mathcal{K}_{Ls}(E)
\left[\rule{0cm}{0.4cm} f_{Ls}(E)-f_{Rs}(E)\right]\;,\nonumber\\
\end{eqnarray}
where $\mathcal{K}_{Ls}(E)$ follows from Eq.~\eqref{eq:k2} with $c=L$ taken for convenience (the current conservation condition
$I_L+I_R=0$ is always fulfilled). The electric and thermoelectric conductances
in the limit of linear 
response are obtained in the same manner
as in Sec.~\ref{sec:cur}. At very low temperature, they read
\begin{subequations}\label{eq:glt02}
\begin{eqnarray}
\frac{G_s}{G_0}&=&\frac{1}{2\pi}\,\mathcal{K}_{Ls}(E_F)\,\mathcal{T}_s(E_F)\equiv g_s(E_F)\,,\label{eq:gt02}\\ 
\frac{L_s}{L_0}&=&\frac{\pi^2 k_BT_0}{3}\,\frac{\partial g_s}{\partial E_F}\,,\label{eq:lt02}
\end{eqnarray}
\end{subequations}

\begin{figure}[t]
\centering
\includegraphics[width=0.45\textwidth]{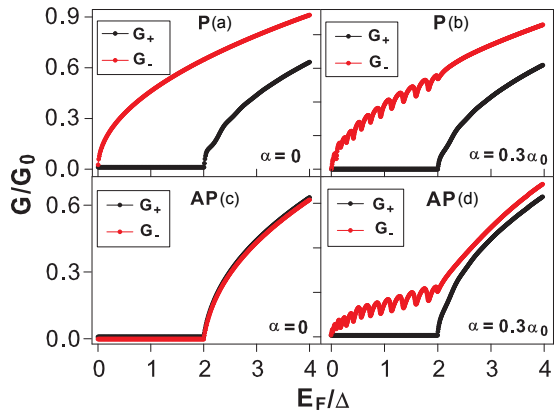}
\caption{(Color online) Spin components of the electric conductance for parallel, (a) and (b), and antiparallel, (c) and (d), configurations as a function of the Fermi energy. Parameter: $\ell=8\ell_0$.}\label{fig:gpmp}
\end{figure}

In Fig.~\ref{fig:gp} we present the total electric conductance, $G=G_++G_-$, as a function of $E_F$. We distinguish between P (a) and AP (b) polarizations of the magnetic leads. Consider first the case without spin-orbit interaction ($\alpha=0$). For parallel orientation the majority electrons are not scattered back while the minority electrons feel a potential well of depth $2\Delta$. Then, when $E_F<2\Delta$ we only have one propagating spin channel, $s=-$, and the conductance is given simply by the number of open channels, which scales 
$\propto E_F^{1/2}$ as discussed in Sec.~\ref{sec:cur}. For $E_F>2\Delta$ the $s=+$ propagating mode becomes active. Thus, the $v_+$ potential causes the transmission to oscillate with energy. These oscillations are akin to the Ramsauer oscillations in electron scattering~\cite{cah03}. This can be more clearly seen in Fig.~\ref{fig:gpmp}(a), which depicts the behavior separately for $G_+$ and $G_-$. The majority spins contributes to $G$ with a $E_F^{1/2}$-dependent function while the minority spins do not attain a nonzero $G$ until $E_F>2\Delta$. Above this energy threshold, the electrons with spin $s=+$ lead to smooth oscillations of $G$ due to the Ramsauer effect. In the antiparallel case [Fig.~\ref{fig:gp}(b)], the conductance is zero below the energy threshold
$E_F=2\Delta$ for both spin indices, see Fig.~\ref{fig:gpmp}(c), since $s=-$ electrons are reflected back from the step potential at the right junction according to Eq.~\eqref{eq:vs} while electronic channels with $s=+$ are not active for $E_F<2\Delta$
due to the step potential at the left contact.

\begin{figure}[t]
\centering
\includegraphics[width=0.45\textwidth]{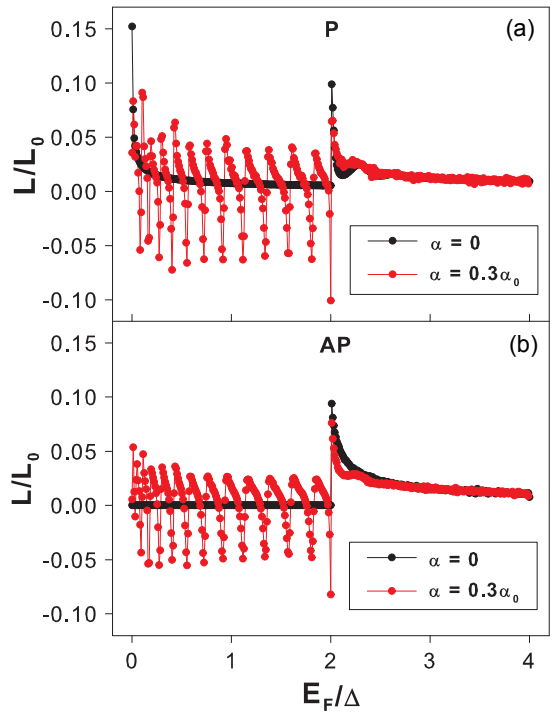}
\caption{(Color online) Thermoelectric conductance as a function of the Femi energy for P (a) and AP (b) polarizations along the $x$-direction. Parameters: $\ell=8\ell_0$ and $k_BT_0=0.01E_0$.}\label{fig:lp}
\end{figure}

Turning on the spin-orbit potential alters the previous picture. For energies below the threshold, Fig.~\ref{fig:gp} shows a different type of oscillation independently of the leads' magnetic orientation. These are due to resonant Fano interference between  the propagating spins and the quasibound states of  opposite spins trapped between the polarized contacts \cite{san06,gel11}.
Figure~\ref{fig:gpmp}(b) shows that the Fano oscillations in the P case are only possible for electrons with spin $s=-$  since for $E_F<2\Delta$ the electrons $s=+$ are evanescent states. At higher energies the two modes become propagating states and the Fano oscillations vanish. In the AP case, the oscillations also appear for $E_F<2\Delta$ but these are now due to electrons with $s=-$  undergoing multiple reflections between the junctions since minority electrons are inactive for transport until $E_F=2\Delta$ \cite{gel11}.

\begin{figure}[t]
\centering
\includegraphics[width=0.45\textwidth]{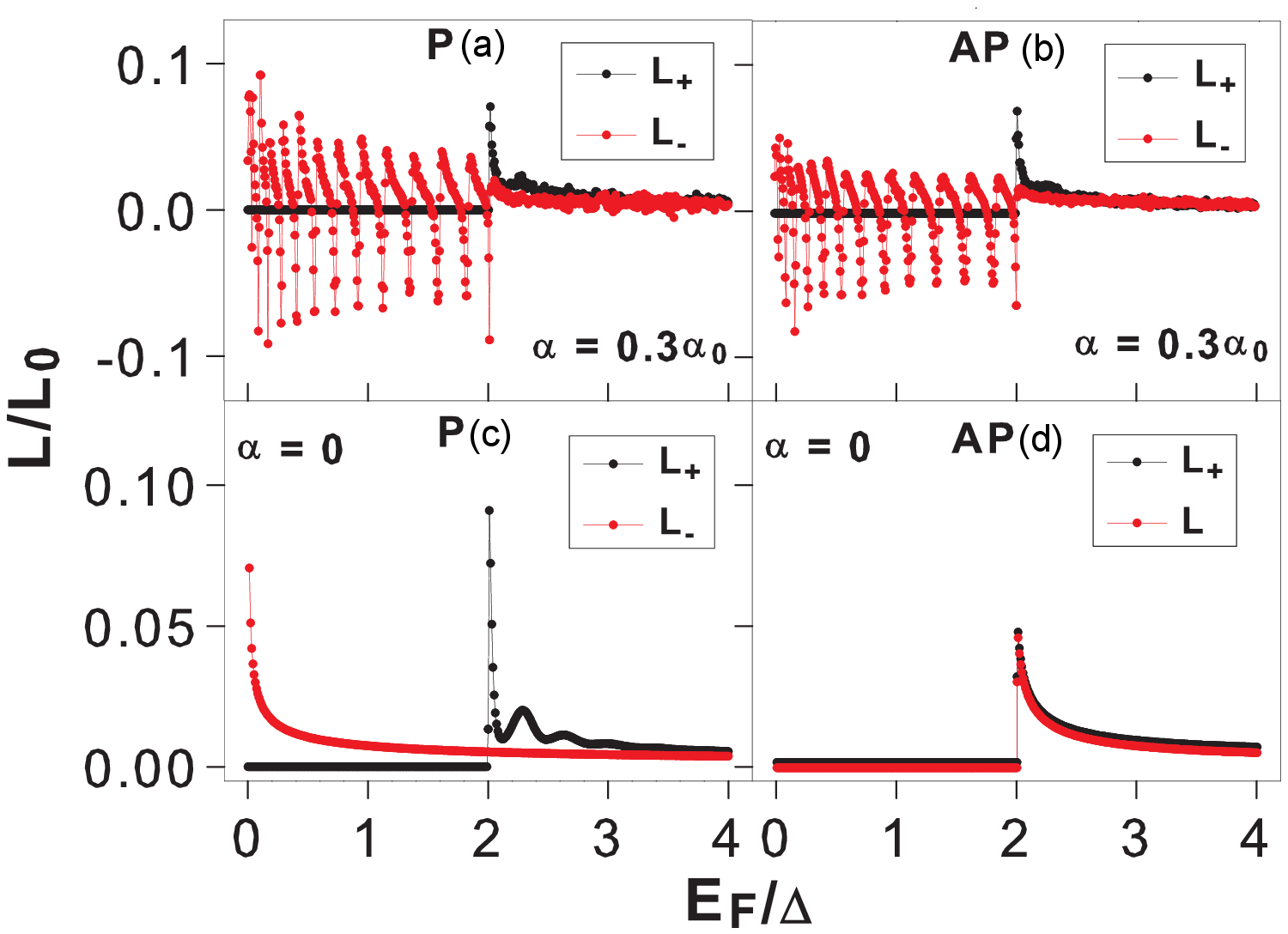}
\caption{(Color online) Spin components of the thermoelectric conductance for parallel, (a) and (c), and antiparallel, (b) and (d), configuration as a function of the Fermi energy. Parameters: $\ell=8\ell_0$ and $k_BT_0=0.01E_0$.}\label{fig:lpmp} 
\end{figure}

Figure~\ref{fig:lp} shows the thermoelectric conductance as a function of Fermi energy for both P and AP configurations. According to Eq.~\eqref{eq:lt02}, $L$ can be written as a function of the energy derivative of the electric conductance. As a consequence, the Rashba induced oscillations for $\alpha\neq 0$ become largely amplified for $E_F<2\Delta$.
Remarkably, the $L$ curves cross the horizontal axis, taking positive and negative values, whereas the case $\alpha=0$
always gives $L>0$. Therefore, the combined influence of ferromagnetic leads and spin-orbit interaction can
drive the electronic current either from the hot to the cold reservoir (as in the normal case, see Fig.~\ref{fig:gllap})
or, notably, from the cold to the hot reservoir. The latter phenomenon is independent of the relative magnetic orientation
(parallel or antiparallel), the main difference being that below the energy threshold the thermoelectric
conductance vanishes in the AP case $\alpha=0$, similarly to the electric conductance [Fig.~\ref{fig:gp}(b)].
Finally, for high energies $L$ smoothly decays to zero since the transmission becomes weakly energy dependent
for $E_F\gg \Delta$. This demonstrates that the spin-orbit interaction in 2D spin transistors leads
to stronger effects for energies lower than the Zeeman splitting.

We depict in Fig.~\ref{fig:lpmp} the spin-resolved thermoelectric conductances. As expected, the large amplitude oscillations in the P case arise in the majority spin channel only [Fig.~\ref{fig:lpmp} (a)]. For $\alpha=0$ the Ramsauer-like oscillations are visible above the energy threshold in the minority channel  [Fig.~\ref{fig:lpmp} (c)].  In the antiparallel case, the thermoconductances obey $L_+=L_-$ and attain their highest value
when $E_F$ reaches $\Delta$ because the transmission variation is largest at that point, as both spin channels become propagating [Fig.~\ref{fig:lpmp} (d)]. In Fig.~\ref{fig:lpmp} (b) the Ramsauer-like oscillations are also visible for $E_F>2\Delta$.
 
\begin{figure}[t]
\centering
\includegraphics[width=0.44\textwidth]{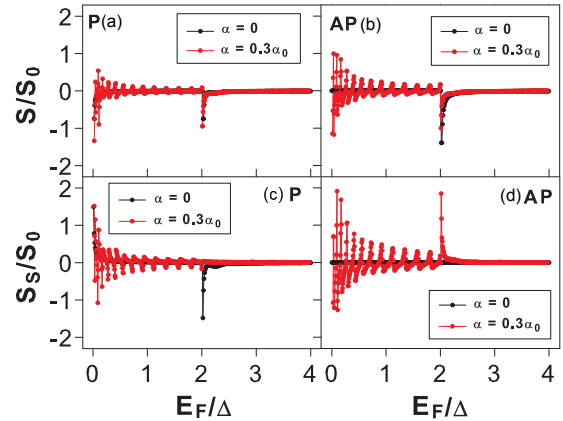}
\caption{(Color online) Seebeck, (a) and (b), and spin-Seebeck, (c) and (d), coefficient as a function of the Fermi energy for both P and AP configurations of leads' magnetic moments pointing along the $x$ direction. Red and black curves represent the cases with and without Rashba coupling in the central region, respectively. Parameters: $\ell=8\ell_0$ and $k_BT_0=0.01E_0$.} \label{fig:sssp}
\end{figure}

The thermopower obeys Eq.~\eqref{eq:sg} but now $g_s$ is calculated from Eq.~\eqref{eq:gt02}. In Fig.~\ref{fig:sssp} we plot the charge Seebec coefficient $S$, for both P (a) and AP (b) configurations. For $\alpha=0$, $S$ is almost zero. Only for energies slightly higher than $2\Delta$ we observe a dip that correlates with the thermoelectric conductance peak observed in Fig.~\ref{fig:lp}. In contrast to the normal case depicted in Fig.~\ref{fig:slap}, here $S$ oscillates and changes its sign for $E_F<2\Delta$. This indicates that at given temperature difference, depending on the value of $E_F$ we can generate positive or negative thermovoltages. This is a very interesting effect since without ferromagnetic contacts the thermopower is always negative. We need to introduce both a Zeeman spliting and a spin-orbit interaction to generate the sign oscillations in $S$, which are more intense for AP configurations. 
We emphasize that although thermopower sign changes can be detected in quantum dots~\cite{sta93,fah13}
or molecular transistors~\cite{red07}, the effect discussed here occurs in an \textit{extended} 2D  system.

To determine the spin Seebeck coefficient $S_{\mathcal{S}}$ at low temperature, we introduce Eq.~\eqref{eq:glt02} in Eq.~\eqref{eq:ss}, obtaining
\begin{eqnarray}
\frac{S_{\mathcal{S}}}{S_0} &=& -\frac{\pi^2k_BT_0}{3}\left(\frac{\partial \ln g_+ (E_F)}{\partial E_F}-\frac{\partial \ln g_- (E_F)}{\partial E_F}\right)\; ,\nonumber\\
\end{eqnarray}
where $g_s(E_F)=\mathcal{K}_{Ls}(E_F)\mathcal{T}_s(E_F)/2\pi$. Here, the spin indices $\pm$ follow the quantization axis established by the leads' magnetization, 
i.e., the $x$-direction.
Unlike the normal case, we now expect nonzero values of $S_{\mathcal{S}}$ since, quite generally,
$g_+\neq g_-$.
Figure~\ref{fig:sssp}(c) shows the results for the parallel configuration in the case with (red curve) and without (black curve) Rashba interaction. We observe that for $\alpha=0$ the spin Seebeck coefficient presents a smooth behavior and only at energies slightly larger than $2\Delta$ the Ramsauer-like oscillations arise. 

When the Rashba coupling is active, $S_{\mathcal{S}}$ changes its sign alternatively for $E_F<2\Delta$. This indicates that at fixed value of $\Delta T$ depending on $E_F$ we can generate a positive or negative spin voltage $V_{\mathcal{S}}$. The case for the AP configuration is plotted in Fig.~\ref{fig:sssp}(d). For zero Rashba coupling $S_{\mathcal{S}}=0$ since the spin components of the electric and thermoelectric conductance are equivalent, see Figs.~\ref{fig:gpmp}(c) and~\ref{fig:lpmp}(d). 
With nonzero value of $\alpha$ we recover the oscillations for $E_F<2\Delta$, with an amplitude larger than for the charge case [Figure~\ref{fig:sssp}(b)].
The oscillations in $S_{\cal S}$ are
more intense for the AP than for the P configuration. It is due to a spin valve effect, i.e., for $E_F<2\Delta$ we do not find any active channel for the AP configuration, while for the P configuration the mode $s=-$ is open. Hence, for $E_F<2\Delta$ 
the electric conductance takes lower values for the AP configuration than for the parallel one.
As a consequence, $S_{\mathcal{S}}$, which is inversely proportional to $G$, attains higher values.

\begin{figure}[t]
\centering
\includegraphics[width=0.44\textwidth]{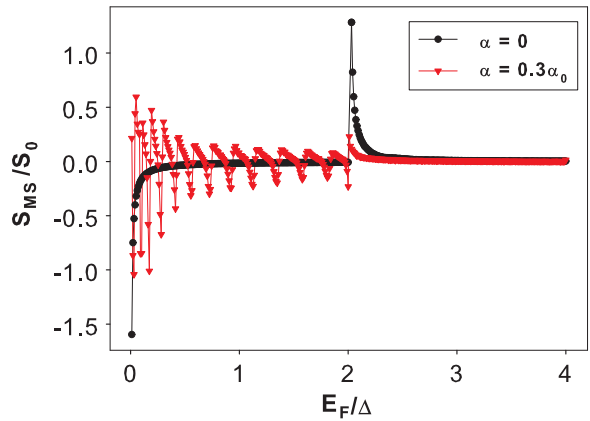}
\caption{(Color online) Magneto-Seebeck coefficient as a function of the Fermi energy at very low temperature. The red and black curves represent the cases with and without spin-orbit interaction in the central region, respectively. Parameters: $\ell=8\ell_0$ and $k_BT_0=0.01E_0$.}\label{fig:smsp}
\end{figure}

The magneto-Seebeck effect is a spintronic phenomenon that gives rise to changes in the thermopower
of a magnetic junction upon switching the leads' magnetic moments \cite{wal11}.
In Fig.~\ref{fig:sssp}(a) and (b) we found that the thermopower significantly changes during the transition from P to AP configurations. Then, it is natural to quantify this departure by defining the magneto-Seebeck coefficient $S_{MS}$ as 
\begin{eqnarray}
S_{MS}=S_P-S_{AP}\,,
\end{eqnarray}
where $S_P$ and $S_{AP}$ are the thermopower in the parallel and antiparallel configuration, respectively.
Figure~\ref{fig:smsp} shows the magneto-Seebeck coefficient as a function of the Fermi energy for $\alpha=0$ (black curve) and $\alpha\neq0$ (red curve). When the Rashba coupling is absent, $S_{MS}$ is closer to zero and only for energies 
slightly exceeding
$2\Delta$ a peak arises since the two spin channels are open and we have an increase of the electric conductance. 
Conspicuously,
for a finite value of $\alpha$ the magneto-Seebeck coefficient presents 
oscillations and sign changes, much like the thermopower oscillations discussed above. This supports the suggestion that the charge thermopower can be controlled by changing the relative magnetization orientation. While the experiment reports sign changes of $S_{MS}$ as a function of the base temperature \cite{wal11}, we here predict a similar effect by tuning the Fermi energy of the 2DES device.

\section{Conclusions}
We have discussed the quantum thermoelectric properties of a two-dimensional electron system with nonhomogeneous spin-orbit interaction. When the device is attached to normal electrodes, we find that the thermopower is strongly modulated by either the 
spin-orbit strength or the channel length at fixed Fermi energy. In the case of ferromagnetic leads,
we distinguish between charge, spin and magneto-Seebeck effects. Interestingly, the thermoelectric dynamics is dominated
by quantum interference effects for energies below the Zeeman splitting in the leads.
These effects lead to large amplitude oscillations of the different thermopowers that change their sign as the Fermi
energy increases. The number of observed oscillations
can be tuned with the separation between contacts. Importantly, the Seebeck coefficients depend on the relative orientation of the magnetic moments in the leads, causing sizeable values of the magneto-Seebeck coefficient.
In general, we demonstrate that a semiconductor two-dimensional electron system offers quite remarkable capabilities
for the generation of highly tunable thermoelectric properties.

Our results may be also relevant for spin transistors built with
two-dimensional electron systems other than semiconductor heterostructures: silicon \cite{app07,den03},
graphene \cite{cho07,sem07,lee12,alo14,dan14} or metal dichalcogenides \cite{dan14}.
Further extensions of our model should consider the quasi-one-dimensional case, which is important for quantum
wires and carbon nanotubes. Another crucial aspect in modeling realistic ferromagnetic-tunnel junctions is the presence
of tunnel barriers, which might alter the predictions discussed in this paper. Future works should also consider
disorder effects, which are relevant in 2D systems with low mobility \cite{and82}, and nonlinear features
as those observed in the dc resistivity of quantum Hall conductors \cite{yan02}.

\section*{Acknowledgments}

This work has been supported by MINECO under Grant No.~FIS2011-23526.

\appendix

\section{Interfaces}\label{app:f}

This appendix contains the mathematical expressions of $\alpha(x)$ and $\Delta(x)$ employed in our numerical simulations for the interfaces.  We model the step-like character of these quantities using Fermi-like functions
\begin{eqnarray}
\mathcal{F}_{x_0,\sigma}(x)=\frac{1}{1+e^{(x-x_0)/\sigma}}\,,
\end{eqnarray}
where $x_0$ is the junction position of the step and $\sigma$ determines the length around $x_0$ where the transition takes place. Then,
\begin{eqnarray}
\alpha(x)&=&\alpha\left(\mathcal{F}_{\ell/2,\sigma_{\alpha}}(x)-\mathcal{F}_{-\ell/2,\sigma_{\alpha}}(x)\right)\,,\\
\Delta(x)&=&\Delta_L \mathcal{F}_{-d/2,\sigma_{\Delta}}(x)+\Delta_R\left(1-\mathcal{F}_{d/2,\sigma_{\Delta}}(x)\right)\,,\nonumber\\
\end{eqnarray}
where the Rashba strength $\alpha$ and the Zeeman splitting $\Delta_{R,L}$ are constants. Our results are independent of the interface details because we take sharp transitions, i.e., $\sigma_{\alpha}\ll\ell$. In our numerical calculations we take $\sigma_{\alpha}=0.1 \ell_0$, $\sigma_{\Delta}=0.3 \ell_0$ and $d=20\ell_0$.

\section{Current expression}\label{app:i}

We obtain the current flowing along $x$-direction from:
\begin{eqnarray}
I_s=\frac{e}{h} \sum_{s^{\prime}}\sum_{q} \!\!\int_{0}^{\infty}\!\!\!\!\! dE \,T_{s^{\prime}s}(E,q) \,[f_{Ls}(E)\!\!-\!\!f_{Rs}(E)] \,,   
\end{eqnarray}
where  $\sum_{s^{\prime}}$ is a sum over the transmitted spin and $\sum_{q}$ a sum over all the possibles momenta on the $y$-direction. Here $f_{cs}(E)$ with $c=L,R$
are the Fermi-Dirac distribution functions of electrons in each contact with spin $s$ [Eq.~\eqref{eq:fermi}] and $T_{s^{\prime}s}$ the transmission probability from subband with spin $s$ to subband with spin $s^{\prime}$.

In the continuum limit we make the replacement:
\begin{equation}
\sum_{q} \longrightarrow  \frac{W}{2 \pi}  \int^{\infty}_{-\infty} dq\, ,
\end{equation}
where $W$ is the width of the sample  in the $y$-direction and $q=\mathcal{K}(E)\sin \theta$ with $\mathcal{K}(E)$ the momentum magnitude and  $\theta$ the incident angle. Using this relation,
\begin{eqnarray}
I_s&=&\frac{e }{h }\frac{W}{2 \pi}\int_{0}^{\infty}\!\!\!\!\!\!dE \int_{-\pi/2}^{\pi/2}\sum_{s^{\prime}} T_{s^{\prime}s}(E,\theta) \,\cos\theta  \,d\theta \\ \nonumber &&\quad\quad\quad\quad\quad \times \,\mathcal{K}(E)\,
[ f_{Ls}(E)\!\!-\!\!f_{Rs}(E)]\,.
\end{eqnarray}

Using Eq.~\eqref{eq:ts} and the fact that $ T_{s^{\prime}s}(E,\theta) \,\cos\theta$ is an even function, we find Eq.~\eqref{eq:Is}.
\newline
\section{Heat transport}\label{app:q}

The stationary heat current (say, at the right contact) is given by:
\begin{eqnarray}\label{eq:q}
Q_R=\frac{1}{h}\frac{W}{2\pi}\sum_s\int_{0}^{\infty}dE\, && \mathcal{K}(E)\,(E-\mu_{Rs}) \,\mathcal{T}_s(E) \nonumber\\
&& \times \left[f_{Ls}(E)-f_{Rs}(E)\right]\,.
\end{eqnarray}
where $\mu_{Rs}$ is the chemical potential in the right contact at given spin $s$. The sum of the right and the left heat current gives the dissipated Joule heating, $Q_R+Q_L=IV$.

We apply a small voltage bias which, after Taylor expanding up to first order in $V$, gives the electrothermal conductance $M=(dQ/dV)_{V=0}$
\begin{eqnarray}\label{eq:m}
M=\frac{e \pi W(k_B T_0)^2}{6 h E_F}\mathcal{K}_F\sum_s\left(\frac{1}{2}\mathcal{T}_s+E_F\frac{\partial \mathcal{T}_s}{\partial E_F}\right)\,
\end{eqnarray}
at very low values of $T_0$. Here, $\mathcal{T}_s=\mathcal{T}_s(E_F)$. Comparing with Eq.~\eqref{eq:lt0}, we check that
the Klein-Onsager relation, $M=T_0L$, is fulfilled as expected.

The thermal conductance, $K=(dQ/d(\Delta T))_{\Delta T=0}$, is obtained in linear response when a small temperature difference is applied across the junction. At low temperatures we find:
\begin{eqnarray}\label{eq:Kc}
K=\frac{\pi W k_B^2 T_0 }{6 h }\mathcal{K}_F\,\mathcal{T}(E_F)\,,
\end{eqnarray}
where $\mathcal{T}(E_F)=\sum_s\mathcal{T}_s(E_F)$. Then, the Wiedemann-Franz law is satisfied since $K/T_0 G=\pi^2 k_B^2/3e^2$.


\begin{thebibliography}{99}
\bibitem{fab07}
 J. Fabian, A. Matos-Abiague, C. Ertler, P. Stano, and I. Zutic, Acta Phys. Slov. \textbf{57}, 565 (2007).
\bibitem{joh85}
M. Johnson and R.H. Silsbee, Phys. Rev. Lett. \textbf{55}, 1790 (1985).
\bibitem{ras60}
E. I. Rashba, Fiz. Tverd. Tela (Leningrad) \textbf{2}, 1224 (1960) [Sov. Phys. Solid State \textbf{2}, 1109 (1960)].
\bibitem{nit97}
 J. Nitta, T. Akazaki, H. Takayanagi, and T. Enoki, Phys. Rev. Lett. \textbf{78}, 1335 (1997).
\bibitem{eng97}
G. Engels, J. Lange, Th. Sch\"apers, and H. L\"uth, Phys. Rev. B \textbf{55}, R1958 (1997). 
\bibitem{gru00}
D. Grundler, Phys. Rev. Lett. \textbf{84}, 6074 (2000).        
\bibitem{dat90}
S. Datta and B. Das, Appl. Phys. Lett. \textbf{56}, 665 (1990). 
\bibitem{koo09}
H. C. Koo, J. H. Kwon, J. Eom, J. Chang, S. H. Han, and M. Johnson, Science \textbf{325}, 1515 (2009). 
\bibitem{jed02}
F. J. Jedema, H. B. Heersche, A. T. Filip, J. J. A. Baselmans, and B. J. van Wees,
Nature \textbf{416}, 713 (2002).
\bibitem{gel10}
M. M. Gelabert, L. Serra, D. S\'anchez, and R. L\'opez, Phys. Rev. B \textbf{81}, 165317 (2010).
\bibitem{zai11}
A. N. M. Zainuddin, S. Hong, L. Siddiqui, S. Srinivasan, and S. Datta,
Phys. Rev. B \textbf{84}, 165306 (2011).
\bibitem{mir01}
F. Mireles and G. Kirczenow, Phys. Rev. B \textbf{64}, 024426 (2001).
\bibitem{san06}
D. S\'anchez and L. Serra, Phys. Rev. B \textbf{74}, 153313 (2006).
\bibitem{jeo06}
J.-S. Jeong and H.-W. Lee, Phys. Rev. B \textbf{74}, 195311 (2006).
\bibitem{zha06}
L. Zhang, F. Zhai, and H. Q. Xu, Phys. Rev. B \textbf{74}, 195332 (2006).
\bibitem{per07}
C. A. Perroni, D. Bercioux, V. Marigliano Ramaglia, and V. Cataudella, J. Phys.: Condens. Matter \textbf{19}, 186227 (2007).
\bibitem{she08}
K. Shen and M.W. Wu, Phys. Rev. B \textbf{77}, 193305 (2008).
\bibitem{san08}
D. S\'anchez, Ll. Serra, and M.-S. Choi, Phys. Rev. B \textbf{77}, 035315 (2008).
\bibitem{ore08}
P. A. Orellana, M. Amado, and F. Dom\'{\i}nguez-Adame,
Nanotechnology \textbf{19}, 195401 (2008).
\bibitem{nag14}
K. E. Nagaev and A. S. Goremykina, Phys. Rev. B \textbf{89}, 035436 (2014).
\bibitem{ser05}
L. Serra, D. S\'anchez, and R. L\'opez, Phys. Rev. B \textbf{72}, 235309 (2005).
\bibitem{qua10}
C. H. Quay, T. L. Hughes, J. A. Sulpizio, L. N. Pfeiffer, K. W. Baldwin,
K. W. West, D. Goldhaber-Gordon, and R. de Picciotto,
Nature Phys. \textbf{6}, 336 (2010).
\bibitem{hau01}
W. Ha\"usler, Phys. Rev. B \textbf{63}, 121310 (2001).
\bibitem{gov02}
M. Governale and U. Z\"ulicke,
Phys. Rev. B \textbf{66}, 073311 (2002).
\bibitem{she05}
E. Ya. Sherman, A. Najmaie, and J. E. Sipe, Appl. Phys. Lett. \textbf{86}, 122103 (2005).
\bibitem{erl06}
S. I. Erlingsson, J. C. Egues, and D. Loss, Phys. Status Solidi C \textbf{3}, 4317 (2006).
\bibitem{upa08}
P. Upadhyaya, S. Pramanik, S. Bandyopadhyay, and M. Cahay,
Phys. Rev. B \textbf{77}, 045306 (2008).
\bibitem{ste10}
P. Stefanski, J. Phys.: Condens. Matter \textbf{22}, 505303 (2010).
\bibitem{mal11}
M. Malard, I. Grusha, G. I. Japaridze, and H. Johannesson,
Phys. Rev. B \textbf{84}, 075466 (2011).
\bibitem{woj14}
P. Wojcik, J. Adamowski, B. J. Spisak, and M. Woloszyn,
J. Appl. Phys. \textbf{115}, 104310 (2014).
\bibitem{kis00}
A. A. Kiselev and K. W. Kim, Phys. Rev. B \textbf{61}, 13115 (2000)
\bibitem{she05b}
E. Ya. Sherman and J. Sinova, Phys. Rev. B \textbf{72}, 075318 (2005).
\bibitem{wu10}
M. W. Wu, J. H. Jian, and M. Q. Weng, Phys. Rep. \textbf{493}, 61 (2010).
\bibitem{gel11}
M. M. Gelabert and L. Serra, Eur. Phys. J. B \textbf{79}, 341 (2011).
\bibitem{sch03}
J. Schliemann, J. C. Egues, and D. Loss, Phys. Rev. Lett. \textbf{90}, 146801 (2003).
\bibitem{hal03}
K. C. Hall, W. H. Lau, K. Gundogdu, M. E. Flatt\'e, and T. F. Boggess,
Appl. Phys. Lett. \textbf{83}, 2937 (2003).
\bibitem{fio14}
G. Fiori, F. Bonaccorso, G. Iannaccone, T. Palacios, D. Neumaier,
A. Seabaugh, S. K. Banerjee, and L. Colombo, Nature Nanotech. {\bf 9}, 768 (2014).
\bibitem{aws07}
D. D. Awschalom and M. E. Flatt\'e, Nature Phys. \textbf{3}, 153 (2007).
\bibitem{uch08}
K. Uchida, S. Takahashi, K. Harii, J. Ieda, W. Koshibae, K. Ando, S. Maekawa, and E. Saitoh, Nature (London) \textbf{455}, 778 (2008).
\bibitem{jaw10}
C.M. Jaworski, J. Yang, S. Mack, D. D. Awschalom,
J. P. Heremans, and R. C. Myers, Nature Mater. \textbf{9}, 898 (2010).
\bibitem{jaw12}
C.M. Jaworski, R. C. Myers, E. Johnston-Halperin, and J. P. Heremans,
Nature (London) \textbf{487}, 210 (2012)
\bibitem{bau10}
G.E.W. Bauer, A.H. MacDonald and S. Maekawa,
Solid State Commun. \textbf{150}, 459 (2010).
\bibitem{sai06}
E. Saitoh, M. Ueda, H. Miyajima, and G. Tatara,
Appl. Phys. Lett. \textbf{88}, 182509 (2006).
\bibitem{wal11}
M. Walter, J. Walowski, V. Zbarsky, M. M\"unzenberg, M. Sch\"afers, D. Ebke, G. Reiss, A. Thomas, P. Peretzki,
M. Seibt, J. S. Moodera, M. Czerner, M. Bachmann, and C. Heiliger, Nat. Mater. \textbf{10}, 742 (2011). 
\bibitem{tei13}
 J. M. Teixeira, J. D. Costa, J. Ventura, M. P. Fernandez-Garcia, J. Azevedo, J. P. Araujo, J. B. Sousa,
 P. Wisniowski, S. Cardoso, and P. P. Freitas, Appl. Phys. Lett. \textbf{102}, 212413 (2013).
\bibitem{lop14}
C. L\'opez-Mon\'{\i}s, A. Matos-Abiague, and J. Fabian,
Phys. Rev. B \textbf{89}, 054419 (2014).
\bibitem{boe14}
A. Boehnke, M. Milnikel, M. Walter, V. Zbarsky, C. Franz, M. Czerner, K. Rott, A. Thomas,
C. Heiliger, and M. M\"unzenber, and G. Reiss, arXiv:1405.1064 (preprint).
\bibitem{nar1}
V. Narayan, M. Pepper, J. Griffiths, H. Beere, F. Sfigakis, G. Jones,
D. Ritchie, and A. Ghosh, Phys. Rev. B \textbf{86}, 125406 (2012).
\bibitem{nar2}
V. Narayan, E. Kogan, C. Ford, M. Pepper, M. Kaveh, J. Griffiths, G. Jones, H. Beere,
and D. Ritchie, New J. Phys. \textbf{16}, 085009 (2014).
\bibitem{chi09}
W. E. Chickering, J. P. Eisenstein, and J. L. Reno,
Phys. Rev. Lett. \textbf{103}, 046807 (2009).
\bibitem{gan12}
A. Ganczarczyk, S. Rojek, A. Quindeau, M. Geller, A. Hucht, C. Notthoff, J. K\"onig, A. Lorke, D. Reuter, and A. D. Wieck,
Phys. Rev. B \textbf{86}, 085309 (2012).
\bibitem{ma10}
Z. Ma, Solid State Comm. \textbf{150}, 510 (2010).
\bibitem{wan10}
C. M. Wang and M. Q. Pang, Solid State Comm. \textbf{150}, 1509 (2010).
\bibitem{dyr13}
A. Dyrdal, M. Inglot, V. K. Dugaev, and J. Barna\'s, Phys. Rev. B \textbf{87}, 245309 (2013).
\bibitem{bor13}
J. Borge, C. Gorini, and R. Raimondi, Phys. Rev. B {\bf 87}, 085309 (2013).
\bibitem{igl14}
P. E. Iglesias and J. A. Maytorena, Phys. Rev. B \textbf{89}, 155432 (2014).
\bibitem{len90}
C. S. Lent and D. J. Kirkner, J.\ Appl.\ Phys.\ {\bf 67}, 6353 (1990).
\bibitem{griffiths}
See, e.g., D. J. Griffiths, Introduction to Quantum Mechanics,
2nd ed., Pearson Prentice Hall, Upper Saddle River, New Jersey (2005), p. 82.
\bibitem{Ashcroft}
N. W. Ashcroft and N. D. Mermin, \textit{Solid State Physics}, Saunders College, Philadelphia (1976),
pp. 253–-258.
\bibitem{alo14}
M. I. Alomar and D. S\'anchez, Phys. Rev. B \textbf{89}, 115422 (2014).
\bibitem{mott}
M. Cutler and N. F. Mott, Phys. Rev. \textbf{181}, 1336 (1969).
\bibitem{sch00}
G. Schmidt, D. Ferrand, L. W. Molenkamp, A. T. Filip, and B. J. van Wees, Phys. Rev. B \textbf{62}, R4790 (2000).
\bibitem{ras00}
E. I. Rashba, Phys. Rev. B \textbf{62}, R16267 (2000). 
\bibitem{slob}
A. Slobodskyy, C. Gould, T. Slobodskyy, G. Schmidt, L.W. Molenkamp, D. S\'anchez, Appl. Phys. Lett. {\bf 90}, 122109 (2007).
\bibitem{cah03}
M. Cahay, S. Bandyopadhyay, Phys. Rev. B \textbf{68}, 115316 (2003).
\bibitem{sta93}
A. A. M. Staring, L. W. Molenkamp, B. W. Alphenaar,
H. van Houten, O. J. A. Buyk, M. A. A. Mabesoone, C.
W. J. Beenakker, and C. T. Foxon, Europhys. Lett. \textbf{22},
57 (1993).
\bibitem{fah13}
S. Fahlvik Svensson, E. A. Hoffmann, N. Nakpathomkun, P. M. Wu, H. Q. Xu, H. A. Nilsson, D. S\'anchez,
V. Kashcheyevs and H. Linke, New J. Phys. \textbf{15}, 105011 (2013).
\bibitem{red07}
P. Reddy, S. Y. Jang, R. A. Segalman, and A. Majumdar,
Science \textbf{315}, 1568 (2007).
\bibitem{app07}
I. Appelbaum, B. Huang, and D. J. Monsma,
Nature (London) \textbf{447}, 295 (2007).
\bibitem{den03}
C. L. Dennis, C. Sirisathitkul, G. J. Ensell, J. F. Gregg and S. M. Thompson,
J. Phys. D: Appl. Phys. \textbf{36}, 81 (2003).
\bibitem{cho07}
S. Cho, Y.-F. Chen, and M. S. Fuhrer,
Appl. Phys. Lett. \textbf{91}, 123105 (2007).
\bibitem{sem07}
Y. G. Semenov, K. W. Kim, and J. M. Zavada,
Appl. Phys. Lett. \textbf{91}, 153105 (2007).
\bibitem{lee12}
M.-K. Lee, N.-Y. Lue, C.-K. Wen, and G. Y. Wu,
Phys. Rev. B \textbf{86}, 165411 (2012).
\bibitem{dan14}
A. Dankert, L. Langouche, M. V. Kamalakar, and S. P. Dash,
ACS Nano \textbf{8}, 476 (2014).
\bibitem{and82}
T. Ando, A. Fowler, and F. Stern, Rev. Mod. Phys. \textbf{54}, 437 (1982).
\bibitem{yan02}
C. L. Yang, J. Zhang, R. R. Du, J. A. Simmons, and J. L. Reno, Phys. Rev. Lett. \textbf{89}, 076801 (2002).

\end{thebibliography}
\end{document}